\documentclass[twoside,10pt]{article}
\usepackage{1}

\begin{document}
\title{\Large \bf Latest oscillation results from T2K }
\author{Marat~Khabibullin \\
{\it Institute for Nuclear Research}\\{\it of the Russian Academy of Sciences}\\ 
{(On behalf of the T2K Collaboration)}
}
\date{}
\maketitle
\begin{center}
{\bf Abstract}\\
\medskip
\parbox[t]{10cm}{\footnotesize
The latest oscillation results obtained in the off-axis accelerator
neutrino experiment T2K are presented. In the data sample, corresponding to 1.43$\times$ 10$^{20}$ protons on target,
 6 $\nu_e$ candidate events pass the selection criteria, while the
expected number of background events for $\sin^2{2\theta_{13}}$=0 is 1.5 $\pm$ 0.3 (syst.). 
The probability to observe six or more candidate events due to background is 0.7\%, equivalent to 2.5$\sigma$ significance.

In the $\nu_\mu$-disappearance analysis the obtained atmospheric oscillation parameters are consistent with results from the Super-Kamiokande
and MINOS experiments.
 }
\end{center}
\section{Introduction} \label{s1}
T2K is a second generation long baseline (LBL) accelerator neutrino experiment:
in contrast to the first generation experiments, like K2K, MINOS and OPERA, T2K has neutrino detectors located
slightly offset with respect to the initial proton beam (off-axis angle is 2.5$^\circ$).

Experiment T2K (Tokai-to-Kamioka) is an International Collaboration of about 500 members from 58 institutes of 12 countries.
The source of muon neutrinos and near detectors are located at the Japan Proton Accelerator Research Complex (J-PARC, Tokai Village,
Ibaraki Prefecture, Japan), while as a far detector the well-known Super-Kamiokande (SK) detector located at 295 km is used
(Kamioka, Gifu Prefecture, Japan).

A primary goal of the T2K is a measurement of the only unknown mixing angle $\theta_{13}$ by detecting
the electron neutrinos at the far detector in the initially almost pure muon neutrino beam (``$\nu_e$-appearance'').

A secondary goal is a precision measurement of so-called atmospheric oscillation parameters $(\theta_{23}, \Delta m^2_{23})$
by detecting a deficit of muon neutrinos at the far detector (``$\nu_\mu$-disappearance'').

\section{Physics motivation: neutrino oscillations} \label{s2}
At present it is known that neutrinos are produced and detected in weak interactions as leptons of three flavours:
electron $\nu_e$, muon $\nu_\mu$ and tau $\nu_\tau$ (for a detailed review of neutrino parameters see~\cite{nu-rev}).
Neutrino flavour eigenstates $|\nu_\alpha\rangle$ ($\alpha = e, \mu, \tau$) are not equal to the neutrino mass 
eigenstates $|\nu_i\rangle$ with mass eigenvalues $m_i$ ($i = 1, 2, 3$). 
A conversion from the mass basis to the flavour basis is governed by
the 3$\times$3 unitary matrix $U_{PMNS}$ (Pontecorvo-Maki-Nakagawa-Sakata)\cite{pont68,mns}, which can be parametrized in such a way, that
it only depends on 3 mixing angles and one CP-violating phase: $\theta_{12}, \theta_{23}, \theta_{13}$ and $\delta_{\rm CP}$.

Two of these four parameters are measured in solar/reactor and atmospheric/accelerator experiments, respectively:
$\theta_{12}\approx 34^\circ$ and $\theta_{23}\approx 45^\circ$.
The corresponding mass squared differences, defined as $\Delta m^2_{ij}\equiv m^2_j - m^2_i$, have the following values:
 $\Delta m^2_{12}\approx 7.6\times 10^{-5}$ eV$^2$/c$^4$ and $|\Delta m^2_{23}|\approx 2.4\times 10^{-3}$ eV$^2$/c$^4$.
The sign of the $\Delta m^2_{23}$ is remained undetermined (``mass hierarchy problem'').

The best upper limit for $\theta_{13}$ was obtained in 1999 by the reactor experiment CHOOZ and slightly corrected
in 2010 by the LBL experiment MINOS: $\theta_{13}< 11^\circ$ ($\sin^2{2\theta_{13}} < 0.15$)~\cite{chooz, minos2010}.
If $\theta_{13}$ has non-zero value, then one can study a potential CP-violation in the lepton sector.

The ~``appearance''~ probability $P(\nu_\mu\to\nu_e)$ to observe an electron neutrino at the distance $L$ from the source of the muon neutrinos
with an inital energy $E$ depends on the mixing angles, mass squared differences and $L/E$ ratio: 

\begin{equation}
P(\nu_\mu\to\nu_e)\approx 
\sin^2(2\theta_{13})\sin^2(\theta_{23})\sin^2(\frac{\Delta m^2_{23} L}{4 E_\nu})\label{eq:001}.
\end{equation}

The ~``disappearance''~ probability $P(\nu_\mu\to\nu_\mu)$ in two-flavour oscillation scenario has the following form: 
\begin{equation}
P(\nu_\mu\to\nu_\mu)\approx 
1 - \sin^2(2\theta_{23})\sin^2(\frac{\Delta m^2_{23} L}{4 E_\nu})\label{eq:002}.
\end{equation}

Using the equations (\ref{eq:001}-\ref{eq:002}) and other inputs one can compute the expected number of events
N$_{SK}^{exp}$ and the neutrino energy spectra at the far detector and compare them with observed number of events
N$_{SK}^{obs}$ and the measured energy spectrum. Fitting these two numbers and/or energy spectra it is possible to
get the parameters in question ($\theta$, $\Delta m^2$).

\section{T2K experimental method} \label{s3}
Muon neutrinos in the accelerator experiments are produced as tertiary particles of proton interactions in a special target.
In the T2K~\cite{t2knim} protons are accelerated at J-PARC  in three stages: 1) at LINAC - up to 400 MeV (currently 181 MeV);
2) at Rapid Cycling Synchrotron (RCS) - up to 3 GeV; 3) at Main Ring (MR) - up to 30 GeV, after which the protons are extracted
into the neutrino beamline in 8 bunches per spill (6 before November 2010).

Neutrino beamline consists of two main parts:
primary section, which transports the protons from the MR to a target, and secondary section, where the secondary particles
(pions, kaons etc.) are produced and decayed. The positive pions produced in a graphite target are collected and focused
into the decay volume by three horns. Muon neutrinos are mainly produced in the $\pi^+$-decays: $\pi^+\to\mu^+\nu_\mu$.
Undecayed hadrons and low energy muons ($p_\mu <$ 5 GeV/c) are absorbed by the beam dump, which is followed by the muon monitors
(MUMON) providing the information on the intensity and profile of the high energy muons. In order to check the intensity,
direction, profile and losses of the proton beam the primary section is equipped with many beam monitors.

\begin{figure}[t]
\begin{center}
\includegraphics[width=0.95\textwidth]{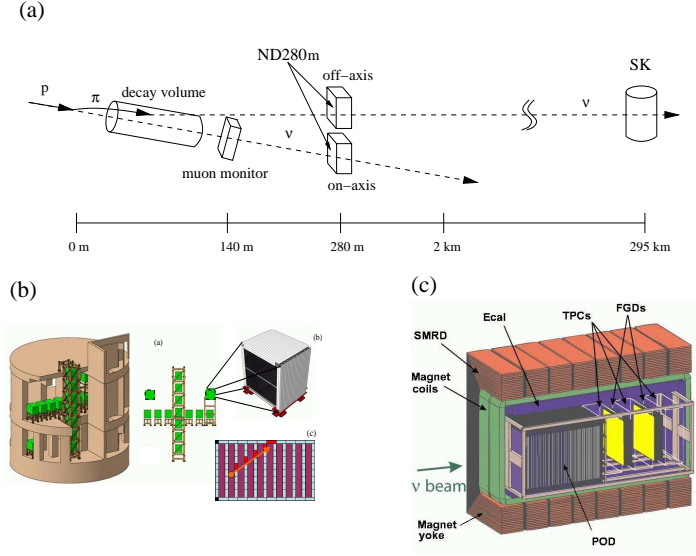}
\end{center}
\vspace{-0.5cm}
\caption{T2K experimental setup: a) a schematic view; b) near detectors: INGRID (left) and the off-axis ND280 (right)}
\vspace{-0.5cm}
\label{fig:1}
\end{figure}

The near detector complex ND280 is located in a specially excavated pit at about 280 meters from the target. It consists of two
independent detectors (Fig.~\ref{fig:1}): the INGRID at $0^\circ$ with respect to the proton beam axis (on-axis), and the ND280 
at 2.5$^\circ$ (off-axis). 
The on-axis near detector INGRID (Interactive Neutrino GRID) is used to monitor the neutrino beam profile,
direction and interaction rates on the day-by-day basis.
The off-axis near detector ND280 consists of a $\pi0$-detector (P0D), a tracker with two fine-grained detectors (FGD) sandwiched by
three time projection chambers (TPC).The tracker and P0D are surrounded by the components of the electromagnetic calorimeter (ECAL),
and all of them are installed inside an UA1/NOMAD magnet which provides a magnetic field of 0.2 T in the direction perpendicular to
the off-axis beam (X-direction). The yoke of the magnet is instrumented as a side muon range detector. The main function of the off-axis ND280
complex is to measure the neutrino flux, energy spectrum, interaction rates and cross-sections {\em before} the oscillation.

The far detector SK is a 50-kton water Cherenkov detector (22.5 kt in the fiducial volume, FV) located at 295 km also at 2.5$^\circ$.
SK detector consists of two main parts: the inner detector (ID) with about 11,100 photomultipliers (20" Hamamatsu PMT) and the outer detector
(OD) with about 1900 PMTs (8"). The main feature of SK detector is an excellent particle identification of muons and electrons with
about 99$\%$ efficiency.

The main advantages of the off-axis conception are as follows: 1) at $0^\circ$ the neutrino
energy is propotional to the parent pion momentum E$_\nu\sim$ p$_\pi$, while at 2.5$^\circ$ the neutrinos have almost monochromatic spectrum with
a high beam intensity; 2) the neutrino energy peak corresponds to the first oscillation maximum; 3) the beam $\nu_e$-contamination at SK is low
($\sim$ 1$\%$);
4) the background from the neutral current (NC) $\nu_\mu$-interactions at the high energy tail is considerably suppressed.

\section{T2K experimental data and selection criteria} \label{s4}
T2K beam data taking was started in January 2010 and paused because of the Great East Japan Earthquake in March 2011.
An analysis of $\nu_e$-appearance~\cite{t2knue} and $\nu_\mu$-disappearance events was carried out for 1.43$\times 10^{20}$
protons on target (p.o.t.)
collected in the first two runs Run I (Jan -- Jun 2010) and Run II (Nov 2010 -- Mar 2011).
The beam power reached 145 kW in March 2011 with 9 $\times$ 10$^{13}$ protons per pulse.

A direction of the off-axis beam during the Runs I and II had being checked by means of the MUMON and INGRID
which demonstrated, that the beam direction was stable well within $\pm$1 mrad (1 mrad shift corresponds to about 2$\%$ shift of the 
E$_\nu$ peak energy at 295 km). INGRID also showed a very stable neutrino interaction rate of about 1.5 events per 10$^{14}$ p.o.t.

The signature of the neutrino interaction in the SK detector is a single electron- or muon-like Cherenkov ring caused by a lepton from a
charged-current quasi-elastic (CCQE) process in the water: $\nu_l + n\to l^- + p$, where $l= e, \mu$. The main backgrounds in case of
the $\nu_e$-appearance are the intrinsic $\nu_e$ from the beam and NC-interactions with $\pi^0\to\gamma\gamma$ in the final states:
$\nu_X + n\to n + \pi^0$ when one photon is missed and another one mimics the electron. In case of the $\nu_\mu$-disappearance the main
background comes from the charged-current processes with one charged pion in the final state (CC1$\pi$): 
$\nu_\mu + n\to\mu^- + n + \pi^+$ or $\nu_\mu + p\to\mu^- + p + \pi^+$.

In order to reject these background events the selection criteria were fixed from Monte Carlo (MC) studies before the data were
collected. The observed number of events N$_{SK}^{obs}$ obtained after applying these selection criteria is compared to the expected
number of events N$_{SK}^{exp}$, computed taking into account a neutrino flux, cross-section predictions and using a normalization factor
from the analysis of events in the off-axis near detector. For the neutrino flux prediction at SK many inputs were
used: the beam monitor data; the hadron production calculations based on the results of the NA61/SHINE CERN experiment~\cite{shine}
and FLUKA MC simulations, also the GEANT3 with GCALOR simulations and cross-sections based on models and external measurements.

To satisfy the general selection criteria related to the $\nu_e$-appearance and $\nu_\mu$-disappearance analyses
the event at SK should have the following parameters: its timing is within
the range from $-$2 to 10~$\mu$s around the beam trigger time; it's a fully-contained (FC) event which means that the vertex and the ring are
within the ID, and there is no activity in the OD. 121 events survived these criteria. This number was reduced to 88 after demanding the
energy deposited in the ID to be at least 30 MeV (visible energy $E_{vis}$) and the vertex
to be in the fiducial volume (FCFV) constrained by an inward 2 meter distance from each ID wall. 41 events have a single Cherenkov ring: 
8 $e$-like and 33 $\mu$-like. 

\section{$\nu_e$-appearance results} \label{s5}
\begin{figure}[tb]
\includegraphics[width=0.47\textwidth]{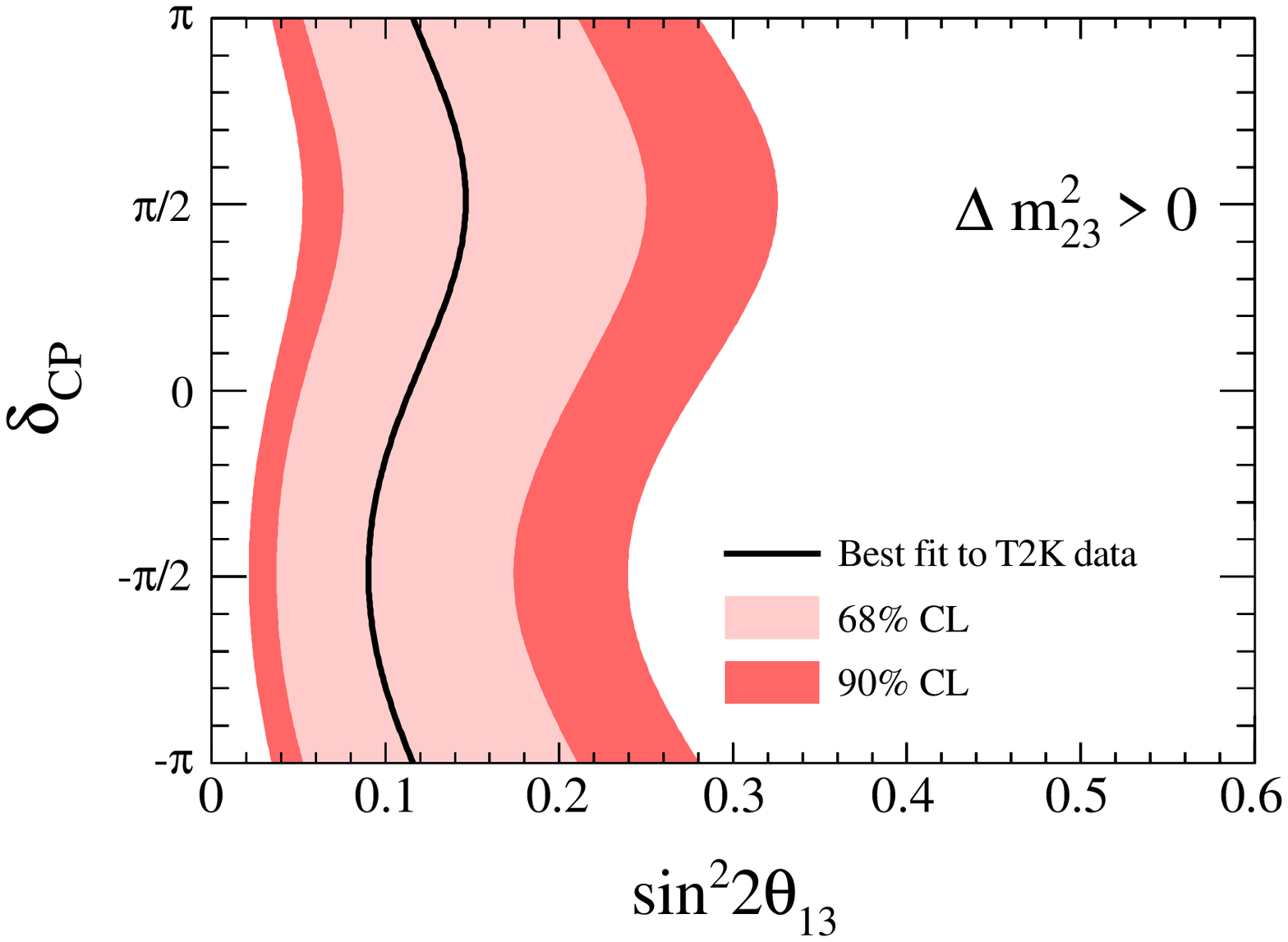}
\includegraphics[width=0.47\textwidth]{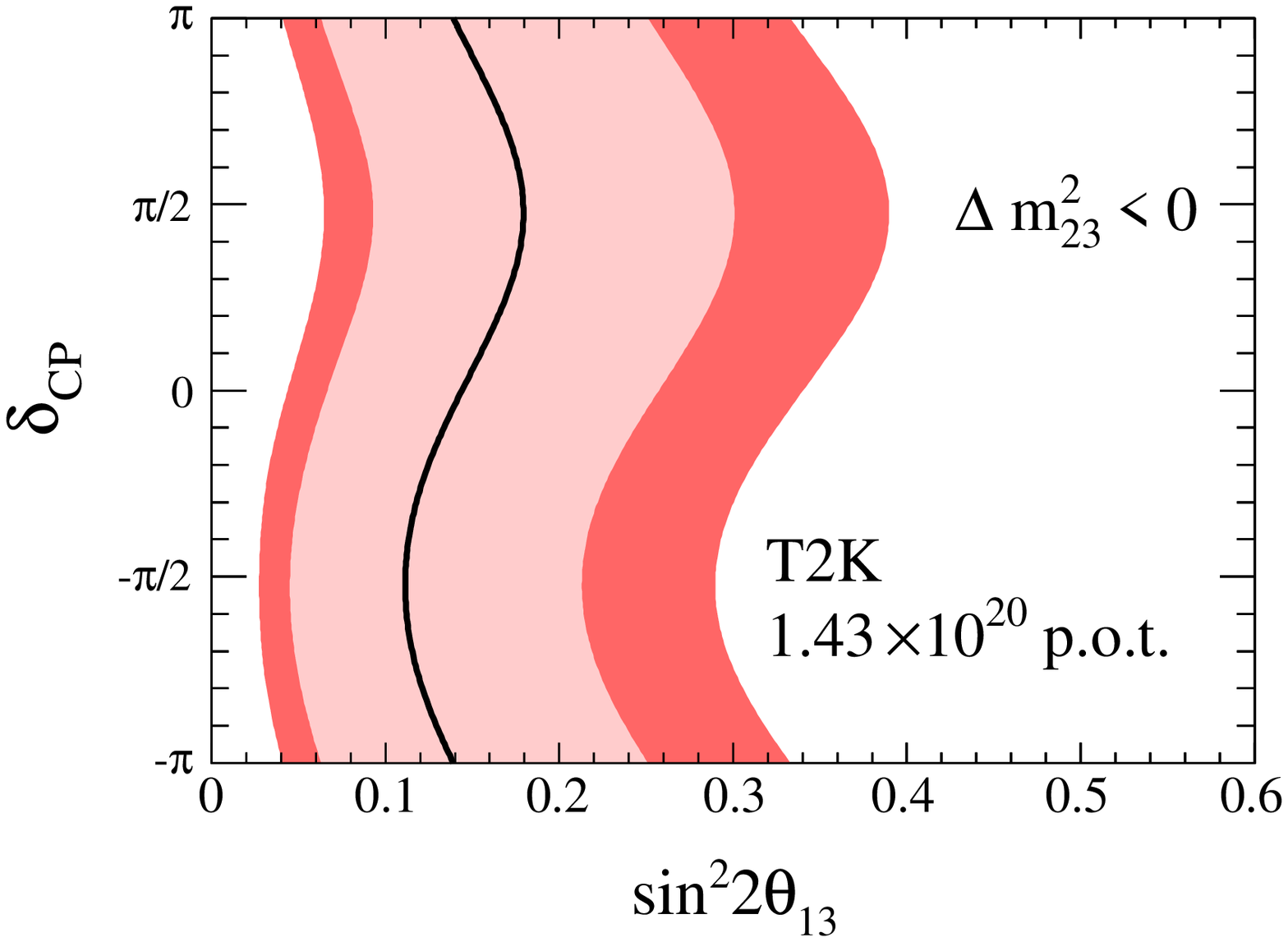}
\vspace{-0.5cm} 
\caption{Contours for $\nu_e$-appearance parameters: the 68\% and 90\%~C.L. regions for $\sin^{2}2\theta_{13}$
for each value of $\delta_{\rm CP}$ for normal (left) and inverted (right) mass hierarchy} 
\vspace{-0.5cm}
\label{fig:2}
\end{figure}
Six out of 8 $e$-like events have $E_{vis}>100$~MeV and no delayed-electron signal.
To suppress misidentified $\pi^0$ mesons, the reconstruction of two
rings is forced, and a cut on the two-ring invariant mass $M_{inv}<105$~MeV$/c^2$ is imposed.
To suppress the background from the intrinsic $\nu_e$ component, the reconstructed neutrino energy required to be 
$E^{rec}_{\nu}<1250$~MeV. No events were rejected after the last two cuts, so, the number of the candidate $\nu_e$-events is $N_{SK}^{obs}$ = 6.
The expected number of events computed for $\sin^2 2\theta_{13}=0$ is $N_{SK}^{exp}$ = 1.5 $\pm$ 0.3, where the total systematic uncertainty
${^{+22.8}_{-22.7}\%}$ is taken into account. 
The probability to observe 6 or more events for $\sin^2 2\theta_{13}=0$ is 0.7$\%$ (2.5$\sigma$ significance).
90\%~confidence intervals for $\theta_{13}$ calculated by the Feldman and Cousins method~\cite{feld-cous} are as follows (Fig.~\ref{fig:2}):
$0.03<\sin^2 2\theta_{13}$ $<$ 0.28 for a normal mass hierarchy ($\Delta m^2_{23}>0$) and $0.04<\sin^2 2\theta_{13}$ $<$ 0.34 for
an inverted mass hierarchy ($\Delta m^2_{23}<0$).

\section{$\nu_\mu$-disappearance results} \label{s6}
\begin{figure}[tb]
\includegraphics[width=0.47\textwidth]{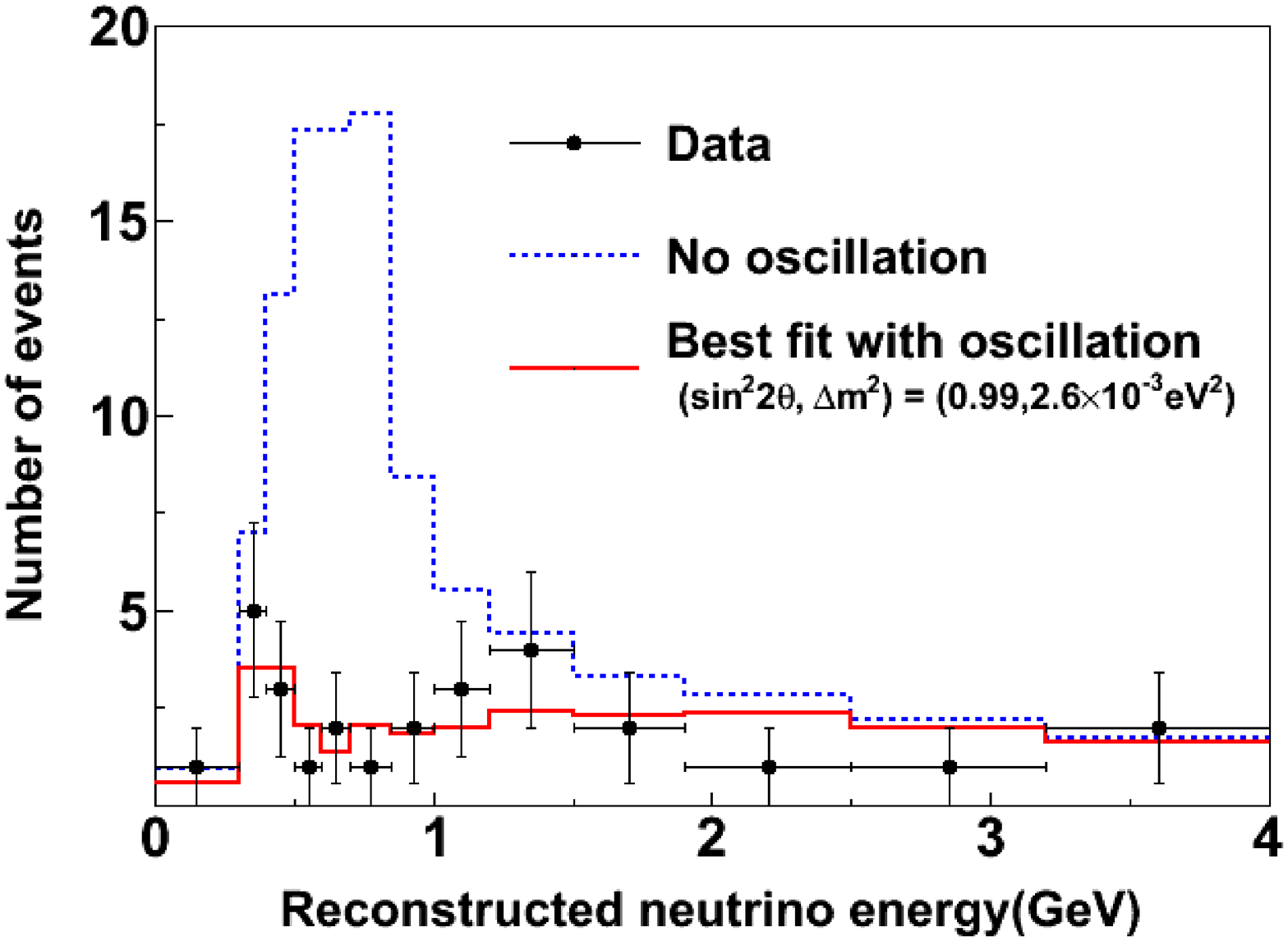}
\includegraphics[width=0.47\textwidth]{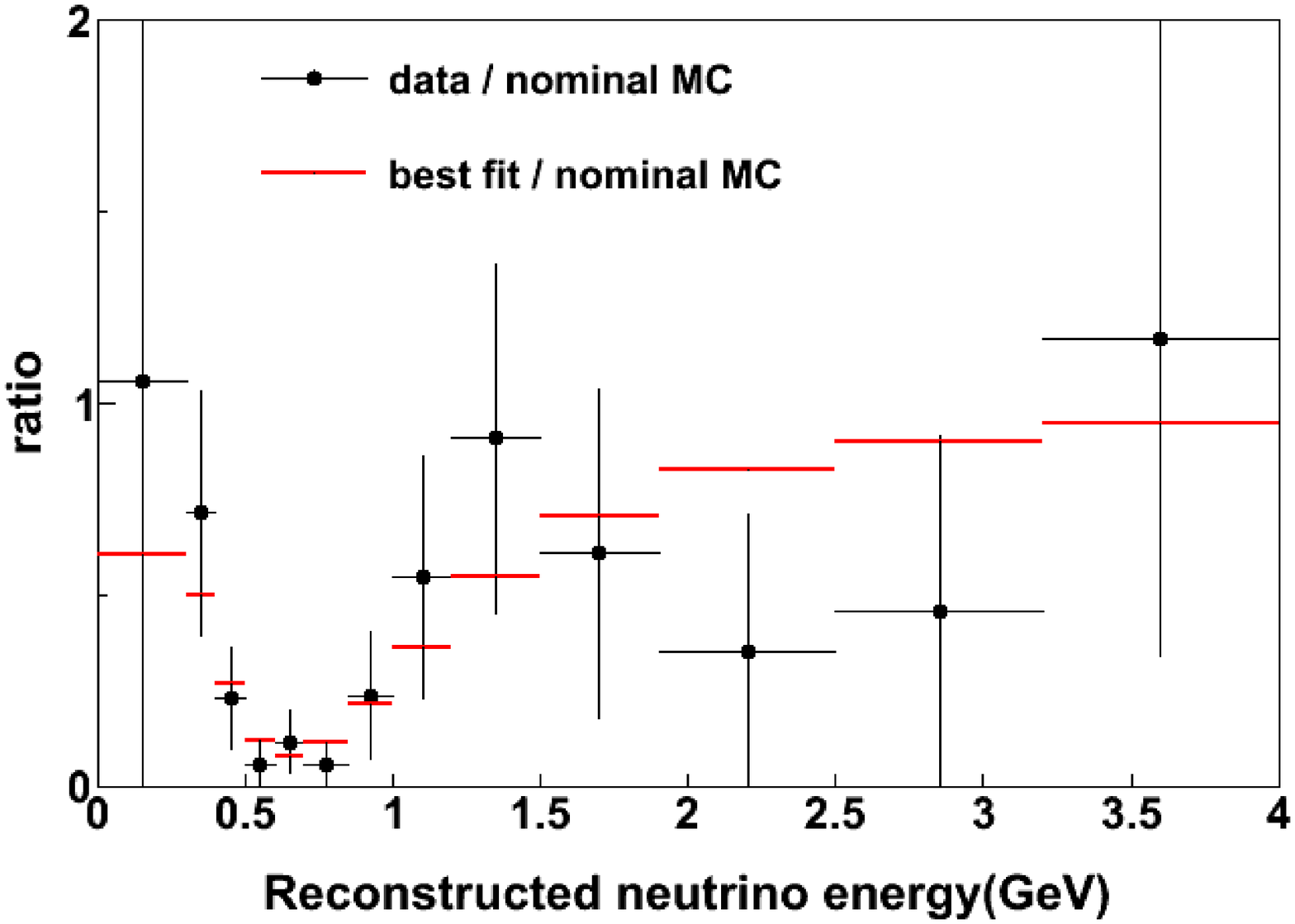}
\vspace{-0.4cm} 
\caption{Reconstructed neutrino energy spectrum at SK (left) and data/MC ratio (right) for $\nu_\mu$-disappearance events} 
\vspace{-0.4cm}
\label{fig:3}
\end{figure}

The 33 events with a single $\mu$-like ring are further checked to reject CC1$\pi$ background events by requiring one or zero
delayed electrons (from the muon decay) and the reconstructed muon momentum p$_\mu >$ 200 MeV/c: 31 events survived.
Under a null oscillation hypothesis the expected number of $\nu_\mu$ is 104 with a systematic uncertainty ${^{+13.2}_{-12.7}}$\%,
which corresponds to a 4.5$\sigma$-significance exclusion of this hypothesis. The reconstructed neutrino energy spectrum at SK demonstrates
a clear oscillation pattern (Fig.~\ref{fig:3}).
The atmospheric oscillation parameters were extracted using two independent fitting methods: A) finding a maximum of a likelihood function
with varying systematic errors, and B) minimizing a special $\chi^2$ with fixed systematic errors.
Both methods gave very close best fit results consistent with the previous measurements by MINOS and SK~\cite{minos-numu,sk-numu}:
$\sin^2{2\theta_{23}}\approx$ 0.99 in the method A (0.98 in method B), $\Delta m^2_{23}$=2.6$\times$10$^{-3}$ eV$^2$/c$^4$.

\section{Conclusions} \label{s7}
The latest oscillation results obtained in the first off-axis accelerator neutrino experiment T2K are presented.
In the data sample, corresponding to 1.43$\times$ 10$^{20}$ p.o.t. (2\% of the final T2K goal), 6 $\nu_e$ candidate
events pass the selection criteria, while the expected number of background events for $\sin^2{2\theta_{13}}$=0 is 1.5 $\pm$ 0.3 (syst.). 
The probability to observe six or more candidate events due to background is 0.7\%, equivalent to 2.5$\sigma$ significance.
At 90\%~C.L., the data are consistent with 0.03(0.04)$<\sin^2 2\theta_{13}<$ 0.28(0.34)  
for $\delta_{\rm CP}=0$ and normal (inverted) hierarchy.

In the $\nu_\mu$-disappearance analysis the obtained atmospheric oscillation parameters are consistent with results from the SK and MINOS
 experiments.

J-PARC plans to restart the work of the accelerator complex in December 2011, and T2K is going to resume the data taking as soon as possible.

This work was supported in part by the "Neutrino Physics" Program of the Russian Academy of Sciences, by the
RFBR (Russia)/JSPS (Japan) grant \textnumero ~11-02-92106 and by the Science School grant \textnumero ~65038.\-2010.2.

\end{document}